\documentclass[a4paper,aps,prd,reprint,superscriptaddress,longbibliography,nofootinbib,floatfix]{revtex4-2}
\usepackage[T1]{fontenc}
\usepackage[utf8]{inputenc}
\usepackage{amsmath}
\usepackage{amssymb}
\allowdisplaybreaks

\usepackage{newtxtext}
\usepackage[smallerops]{newtxmath}
\usepackage[dvipsnames]{xcolor}
\usepackage{graphicx}

\usepackage{float}
\usepackage[italicdiff]{physics}
\usepackage[colorlinks=True,allcolors=blue]{hyperref}
\usepackage{microtype}
\usepackage{orcidlink}

\usepackage{cleveref}
\Crefname{section}{Sec.}{Secs.}
\Crefname{equation}{Eq.}{Eqs.}
\Crefname{figure}{Fig.}{Figs.}
\Crefname{table}{Table}{Tables}

\newcommand\technionCH{Schulich Faculty of Chemistry, Technion - Israel Institute of Technology, Haifa 3200003, Israel}

\newcommand\technionQC{Helen Diller Quantum Center, Technion - Israel Institute of Technology, Haifa 3200003, Israel}

\newcommand\xmumPH{School of Mathematics and Physics, Xiamen University Malaysia, 43900 Sepang, Malaysia}

\newcommand\iitrPH{Department of Physics, Indian Institute of Technology Roorkee, Roorkee 247667, India}

\newcommand\iitrQC{Centre for Photonics and Quantum Communication Technology, Indian Institute of Technology Roorkee, Roorkee 247667, India}

\newcommand\ugPH{Institute of Theoretical Physics and Astrophysics, Faculty of Mathematics, Physics and Informatics, University of Gda\'nsk, 80-308 Gda\'nsk, Poland}
\begin{document}
%%%%%%%%%%%%%%%%%%%%%%%%%%%%%%%%%%%%

%%%%%%%%%%%%%%%%%%%%%%%%%%%%%%%%%%%%
\title{Probing the Strong Equivalence Principle through the External Field Effect. \\ How Do Two Masses Fall?}

\author{Ankit Kumar\,\orcidlink{0000-0003-3639-6468}}
\thanks{These authors contributed equally to this work.}
\affiliation{\technionCH}
\affiliation{\technionQC}

\author{Kelvin Tang Tee Tniam\,\orcidlink{0009-0005-3684-1244}}
\thanks{These authors contributed equally to this work.}
\affiliation{\xmumPH}

\author{Peng Chengxiaohe\,\orcidlink{0009-0004-5540-5520}}
\affiliation{\xmumPH}

\author{Ng Li Yang\,\orcidlink{0009-0005-2940-5241}}
\affiliation{\xmumPH}

\author{P. Arumugam\,\orcidlink{0000-0001-9624-8024}}
\affiliation{\iitrPH}
\affiliation{\iitrQC}

\author{Tom Złośnik\,\orcidlink{0000-0001-7715-5842}}
\affiliation{\ugPH}

\author{Yen-Kheng Lim\,\orcidlink{0000-0002-0907-6904}}
\affiliation{\xmumPH}

\author{Tomasz Paterek\,\orcidlink{0000-0002-8490-3156}}
\email{tomasz@paterek.info}
\affiliation{\xmumPH}
\affiliation{\iitrQC}
\affiliation{\ugPH}

%%%%%%%%%%%%%%%%%%%%%%%%%%%%%%%%%%%
\begin{abstract}
Despite compelling astrophysical evidence for dark matter, the absence of a confirmed dark matter particle has sustained interest in modified gravity as an alternative explanation for the observed phenomenology. One prominent example is Modified Newtonian Dynamics (MOND), which predicts that the internal dynamics of a system depends on the external gravitational field in which it is embedded. This so-called External Field Effect violates the strong equivalence principle and is absent in Newtonian mechanics and general relativity, making it a promising avenue for experimental tests of modified gravity. Motivated by this possibility, we investigate the dynamics of two spherical masses arranged such that their symmetry axis is either parallel or orthogonal to the local gravitational field. Within MOND, we derive solutions describing the internal dynamics of such systems in both strong uniform and radial external fields. In particular, for the case of a radial external field, if the non-relativistic gravitational field is free to have non-vanishing curl, we find that the mutual attraction of the masses in the perpendicular configuration is not strictly aligned with their symmetry axis. It acquires a small transverse component, even when the external gravitational field is everywhere balanced by non-gravitational forces. Using these solutions, we determine the spatial and temporal sensitivities required to distinguish the two configurations and systematically assess experimentally relevant effects, including air drag, object size, and surface interactions. As a representative example, detecting the prediction of the commonly used simple MOND interpolating function requires a spatial sensitivity of order 0.1 fm for sub-millimeter masses evolving over approximately 30 minutes. Such integration times may be achievable with levitated particles or in space-based environments. More broadly, experiments operating at lower resolutions are also interesting as independent tests of the strong equivalence principle and place constraints on modified-gravity theories.
\end{abstract}
%%%%%%%%%%%%%%%%%%%%%%%%%%%%%%%%%%%%

%%%%%%%%%%%%%%%%%%%%%%%%%%%%%%%%%%%%
\maketitle
%%%%%%%%%%%%%%%%%%%%%%%%%%%%%%%%%%%%

%%%%%%%%%%%%%%%%%%%%%%%%%%%%%%%%%%%%
\section{Introduction}
%%%%%%%%%%%%%%%%%%%%%%%%%%%%%%%%%%%
Numerous astrophysical observations suggest the existence of dark matter (DM), including galaxy rotation curves \cite{ComaCluster-Zwicky1937}, gravitational lensing \cite{Allen:1997vc}, and the large-scale structure in the universe \cite{Angulo:2021kes}.
The term ``dark matter'' originates from the hypothesis of an invisible form of matter that accounts for the missing mass in spiral galaxies, where visible luminous matter alone is insufficient to explain the observed stable orbits of stars \cite{ComaCluster-Zwicky1937}.
Despite extensive astronomical and terrestrial searches over several decades, the fundamental nature of DM remains unresolved \cite{DM-Review-Bertone2018}, fueling a continued interest in alternative explanations.

One such alternative is Modified Newtonian Dynamics (MOND), which proposes a modification of the laws of gravity at small Newtonian accelerations,
thereby resolving some of the observed discrepancies without invoking DM candidates \cite{MOND-Milogram1983}.
Although its physical origins remain unclear, MOND has demonstrated notable predictive power across a wide range of phenomena using only a few free parameters. In particular, it successfully fits the rotation curves of various spiral galaxies \cite{MOND-Milogram1983,Milgrom:1983pn,Milgrom:1983zz,MOND-Hees2015}, and naturally predicts the observed baryonic Tully–Fisher relation \cite{Tully:1977fu}, and radial acceleration relation \cite{McGaugh:2016leg}.
The latter two are arguably especially significant as they offer a more direct explanation than within the standard $\Lambda$CDM model \cite{Peebles:1984ge,Bertone:2004pz,Perivolaropoulos:2021jda}.
Furthermore, MOND has been extended to relativistic formulations that are capable of explaining the observed pattern of anisotropies of the cosmic microwave background \cite{RMOND-Zlosnik2021}, and is considered an alternative to the `Planet Nine Hypothesis' in explaining the distribution of extreme trans-Neptunian objects \cite{MOND-P9-Migaszewski2023,MOND-P9-Mathur2023}. 
However, challenges remain. The original MOND model is inherently non-relativistic and so its scope of application is unclear. When applied to explain gravitational lensing and mass discrepancies in clusters of galaxies, it does not adequately account for the data \cite{Famaey:2013ty,Sellwood_2022,Pittordis_2023}. It is an open question as to whether this reflects the breakdown of `the MOND limit' of some of its putative relativistic extension, or reflects a genuine obstacle for any model of gravitation that includes MOND-type phenomenology.
Despite these open issues, MOND's empirical successes motivate terrestrial laboratory tests of this and similar models.

MOND leads to several non-canonical predictions, see e.g. \cite{Lag4MOND-Bekenstein1984,EFE-Blanchet2011,MONDinMilkyWay-Famaey2005,Kumar_MOND_Entanglement}. For example, the theory is nonlinear and does not obey the superposition principle.
Accordingly, even simple mass configurations can produce gravitational fields that differ from Newtonian counterparts. We illustrate these implications in \Cref{APP_MOND}, and here focus on the External Field Effect (EFE), which implies that the internal gravity of a system depends on the external gravitational field that it resides in.
To probe the EFE, we propose an experiment comparing the mutual gravitational attraction of two masses arranged such that their symmetry axis is either parallel or perpendicular to the local gravitational field of the Earth.
Under the assumption of a uniform external field, both Newtonian gravity and general relativity predict identical behavior for the two configurations, i.e., the evolution of separation between freely-falling masses is independent of the orientation of their symmetry axis.
In general relativity, this follows directly from the equivalence principle, which implies that a uniform gravitational field is indistinguishable from an accelerated reference frame. In such a frame, the external field is absent, and only the mutual interaction between the masses remains.
In Newtonian mechanics, the same result follows from the superposition principle. A uniform external acceleration affects both masses equally and can be removed, leaving unaffected the relative dynamics.

In contradistinction, the EFE predicts that the mutual attraction depends on the orientation of the system relative to the external field. 
The actual magnitude of this effect in MOND depends on a specific solution, as will be shown in \Cref{SEC_UNIFORM}.
This anisotropy signals a violation of the strong equivalence principle, though the weak equivalence principle is still honored \cite{Lag4MOND-Bekenstein1984}.
The existence of EFE is also motivated by astronomical observations. For instance, there is evidence that some dwarf galaxies near the Milky Way exhibit no signs of DM, despite residing in low-acceleration regimes where MOND effects should dominate \cite{Klypin:1999uc}.
This is explained by noting that the external gravitational field of the Milky Way restores practically Newtonian internal dynamics through the EFE.
Even if the experimental sensitivity required to detect MOND-specific predictions is not achieved, the proposed setup remains valuable as an independent test of the strong equivalence principle.

The structure of the paper is as follows.
In the following \Cref{sec:no-field} we study the bipartite dynamics in the absence of an external field. We identify constraints on bead size, examine the role of surface interactions, analyze exact trajectories and possibility of avoiding high-vacuum environments.
In \Cref{SEC_STRONG}, we review the EFE in the presence of a strong external field. \Cref{SEC_UNIFORM} focuses on a uniform external field, where we present solutions with both vanishing and non-vanishing curl of the internal field, and derive the spatio-temporal resolution required for an experimental probe. The generalization to a radial external field is presented in \Cref{SEC_RADIAL}. We provide additional details on the MOND framework in \Cref{APP_MOND}, and the Blanchet–Novak solution \cite{Lag4MOND-Bekenstein1984,EFE-Blanchet2011} in \Cref{APP_RESCALING}.

%%%%%%%%%%%%%%%%%%%%%%%%%%%%%%%%%%%%
\section{No external field}
\label{sec:no-field}
%%%%%%%%%%%%%%%%%%%%%%%%%%%%%%%%%%%

\begin{figure}[!t]
    \centering
    \includegraphics[width=\linewidth]{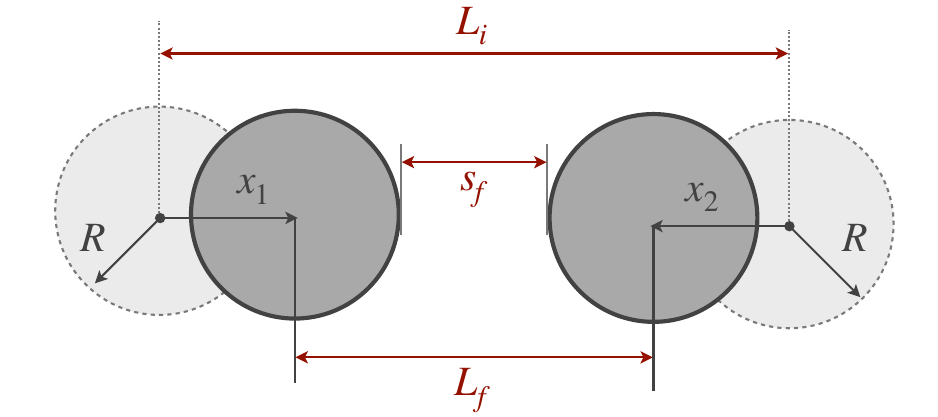}
    \caption{Two identical masses of radius $R$ in the absence of an external field.
    The initial and final (center-to-center) separations are $L_i$ and $L_f$, respectively, and $s_f$ is their final surface-to-surface separation. Displacements from the initial positions are denoted by $x_1$ and $x_2$.}
    \label{fig:setup}
\end{figure}

We begin by establishing the conditions under which the motion of two beads, driven by their small mutual gravitational attraction, can be observed in the absence of an external field.
Consider the configuration shown in \Cref{fig:setup}.
Two masses are initially at rest with center-to-center separation $L_i$, and upon release they attract each other gravitationally, reaching a final separation $L_f$ which may allow for a finite surface-to-surface gap $s_f$. We denote the displacements of the two beads from their initial center-of-mass positions by $x_1$ and $x_2$. By symmetry, $x_2 = - x_1$ at all times.

%%%%%%%%%%%%%%%%%%%%
\subsection{Minimal radius}

To reproduce galaxy rotation curves, MOND modifies the law of gravitation\footnote{In an alternative realization of MOND, the law of gravitation is kept intact, but Newton's second law is modified (so-called modified inertia \cite{Milgrom:2011kx}). However, it suffers from problems related to the absence of a known action-based formulation, e.g., energy conservation \cite{Dreyer2010}. Here we follow the modified gravity approach, as proposed by Bekenstein and Milgrom (AQUAL) \cite{Lag4MOND-Bekenstein1984}.} for objects experiencing Newtonian accelerations $a_0 \lesssim 1.2 \times 10^{-10}$ m/s$^2$ \cite{MOND-Milogram1983,MONDinMilkyWay-Famaey2005,MOND-Gentile2011}.
For larger accelerations, Newtonian gravity is recovered.
The necessary condition for our proposal is therefore that mutual gravity between the beads is $\sim a_0$ throughout the motion.
Using the final separation $L_f$, this condition reads $Gm/L_f^2 < a_0$, where $G$ is Newton's constant and $m$ is the mass of each spherical bead.
An equivalent condition can be obtained using the MOND low-acceleration relation $\sqrt{a_0 a_N}$, since $a_0$ marks the transition between Newtonian and MOND regimes.
Expressing the constraint in terms of the ratio between the surface-to-surface distance and mass diameter yields
\begin{align}
1 + \frac{s_f}{2 R} > \sqrt{G \frac{\pi \rho R}{3 a_0}},
\label{EQ_SR}
\end{align}
where $\rho$ is the material density.
Note that this ratio depends on the bead size: smaller masses can be placed relatively closer together, whereas larger masses must remain relatively farther apart. Requiring the masses to attract each other with acceleration exactly equal to $a_0$ when their surfaces are in contact, i.e. $s_f = 0$, fixes the bead radius to
\begin{align}
R_0 = \frac{3}{\pi} \frac{a_0}{G \rho}.
\label{EQ_R_BOUND}
\end{align}
Beads with radii smaller than $R_0$ experience accelerations below $a_0$ throughout the entire evolution, even when $s_f = 0$, and are therefore suboptimal. 
Thus, $R_0$ is the minimal radius for which the final acceleration can reach $a_0$.
Conversely, for beads with radii larger than $R_0$, the evolution must be terminated at a finite value of $s_f$ in order to keep the acceleration within the MOND regime, i.e. below $a_0$.
In our numerical estimates we assume the beads to be made of Platinum\footnote{Platinum is highly dense, $\approx 21.45 \, \mathrm{g/cm}^3$, and at the same time malleable to form smooth and non-corrosive spheres of predetermined size.}, for which $R_0 \approx 80 \, \mu$m.
As an illustration, \Cref{EQ_SR} requires Platinum spheres of radius $100 \,\mu$m to have a final surface-to-surface separation of at least $\approx 23 \,\mu$m, whereas for spheres of radius $1$ cm the corresponding minimum separation is $\approx 20$ cm. 
%%%%%%%%%%%%%%%%%%%%%%%%%%%%%%%%%%%%%
%%%%%%%%%%%%%%%%%%%%%%%%%%%%%%%%%%%%%

%%%%%%%%%%%%%%%%%%%%%%%%
\subsection{Surface effects}
\label{SEC_VDWAALS}

The quantitative numbers presented in the previous section show that in microscopic experiments non-gravitational surface forces may dominate the dynamics and obscure the gravitational signal.
To ensure that this does not occur, we require the van der Waals acceleration at the final surface-to-surface separation $s_f$ to remain of order $a_0$.
For an estimation, the proximity limit $s_f \ll R$ of the van der Waals acceleration is given by \cite{HAMAKER19371058}:
$A R/ 12 s_f^2 m$,
where $A \approx 4.5 \times 10^{-19} \, \text{J}$ is the Hamaker coefficient for Platinum \cite{TOLIAS2020121652}.
Equating this expression to $a_0$ and using $s_f$ from \Cref{EQ_SR} refines the minimal radius to approximately $\approx 215 \, \mu$m. 
For this radius, corresponding to a final separation $s_f = 274~\mu$m, the exact (non-proximity) calculation gives a van der Waals acceleration approximately twice as large as $a_0$.

The acceleration due to Casimir forces is given by \cite{Casimir-Emig2007}:
\begin{align}
    a_\text{Cas} =
-\dfrac{\hbar c}{m\pi}
{\displaystyle\sum_{n=0}^{\infty}}
C_n  \dfrac{(n+7) R^{n+6}}{(2R+s_f)^{n+8}} , 
\end{align}
for $s_f \sim R$ and with the coefficients $\{C_n\}$ determined numerically in \cite{Casimir-Emig2007}. For Platinum spheres of radius $215\,\mu$m and final surface-to-surface separation $274\,\mu$m, one obtains $a_\text{Cas} \approx 7.8 \times 10^{-15} \, \text{m/s}^2 \sim a_0/15000$, showing that Casimir forces are negligible compared to van der Waals effects.

%%%%%%%%%%%%%%%%%%%%%%%%%%%%%%%%%%%%
\subsection{Duration of the experiment}

The weakness of gravity implies that the expected displacements are small, independently of the size of the beads.
In the first approximation, the masses can be assumed to undergo a uniformly accelerated motion, with the final acceleration $a_0$, providing the upper bound on their displacement as $x_1 < \frac{1}{2} a_0 t^2$.
Thus, increasing the duration by one order of magnitude increases the displacement by two orders of magnitude.
In terrestrial free-fall experiments, the observation time is limited by apparatus size. For example, the Bremen drop tower allows free-fall durations below $10$ seconds for samples that are catapulted upwards \cite{VONKAMPEN2006278}, corresponding to a maximum displacement of order $10$ nm. 
In orbits around the Earth or using suitably levitated masses, the experiment could run for much longer. As an illustration, to which we will return later, for $30$ minutes of evolution, the distance between the particles changes by $\lesssim 0.4$ mm.

We now analyze exact displacements, accounting for changes in mutual acceleration.
From energy conservation in Newtonian mechanics one derives the following equation for time $t$ it takes each mass to move distance $x_1$:
\begin{align}
t =&  \sqrt{ \frac{L_f + 2 x_f}{2 G m}} \Bigg[  \sqrt{x_1 \left( L_f + 2(x_f-x_1) \right) }
\notag\\%%%%%%%%%%%
&\quad +  \frac{L_f + 2 x_f}{2 \sqrt{2}}\left( \frac{\pi}{2} - \arctan( \theta_t) \right) 
\Bigg],
\label{EQ_T_X}
\end{align}
where $\theta_t = ( L_f + 2 x_f - 4 x_1) / \sqrt{8 x_1 (L_f + 2 (x_f - x_1))}$,
and $x_f$ is the final displacement of each mass corresponding to the center-to-center separation $L_f$.
We expressed $t$ in terms of $L_f$ as the latter is fixed by the acceleration requirements, see Ref. \cite{Krisnanda_2020} for an expression in terms of the initial separation $L_i$.
Since \Cref{EQ_T_X} is transcendental in $x_1$, we proceed further by introducing the ratio $\epsilon_f = x_f / L_f$ and $\epsilon_1 = x_1 / L_f$, with $\epsilon_1 < \epsilon_f$.
For many experimentally relevant durations we have $\epsilon_f \ll 1$, which implies $\theta_t \gg 1$, and justifies approximation $\arctan( \theta_t) \simeq \frac{\pi}{2} - \frac{1}{\theta_t}$.
Accordingly, a series expansion reads:
\begin{align}
t \sqrt{ \frac{G m}{2 L_f^3}} & \simeq  \sqrt{\epsilon_1} + \sqrt{\epsilon_1} \epsilon_f + \dots
\label{EQ_T_APPROX}
\end{align}
Note that ignoring the second and further terms in \Cref{EQ_T_APPROX} returns the familiar result $x_1 = G m t^2 / 2 L_f^2 = \frac{1}{2} a_0 t^2$, and therefore those terms quantify contributions from the non-uniform acceleration.

%%%%%%%%%%%%%%%%%%%%%%%%%%%%%%%%%%%%%%%%%%%
\subsection{Air drag}

Since the required experimental parameters indicate a fairly macroscopic setup, we explore if measurements can be performed without high vacuum conditions. 
For small bead velocities the air acts as a viscous fluid, exerting a drag force given by $F_\text{drag} = 6 \pi \eta  R v$, where $\eta \approx 1.8 \times 10^{-5} \,  \text{Pa}\cdot\text{s}$ is the dynamic viscosity coefficient of air at room temperature and pressure, and the velocity $v$ can be upper bounded by the value for the uniformly accelerated motion $v \approx a_0 t$. The deceleration due to air friction is therefore $a_{\text{drag}} = F_\text{drag} / m \approx 9 \eta a_0 t / 2 \rho R^2$, and decreases with the size of the object.
For $R = 215 \,\mu$m, even short experiments for $t = 10$ s yield deceleration comparable to $a_0$.
A longer experiment, for $t = 10^4$ s, would require a bead with $R \approx 2$ cm for the drag deceleration to be one order of magnitude smaller than $a_0$.
We therefore conclude that microscopic experiments require high vacuum, and only larger-scale beads can be observed in a regular, well-sealed chamber.

%%%%%%%%%%%%%%%%%%%%%%%%%%%%%%%%%%%%%%%%%%%
\section{External field effect}
\label{SEC_STRONG}
%%%%%%%%%%%%%%%%%%%%%%%%%%%%%%%%%%%%%%%%%%%

Consider an arbitrary mass distribution embedded in a strong external gravitational field. This situation has previously been studied in Ref. \cite{Lag4MOND-Bekenstein1984} through the solution for the potential.
Here, we adopt a formulation in terms of the gravitational field as it turns out that solutions  exist where fields do not admit a scalar potential. Although it should be emphasized that existing action-based formulations of MOND do lead to the gravitational field being the gradient of one of more scalar potentials, we do not a-priori want to exclude the possibility that the field represents a more complicated object. The starting equation of the MOND model reads:
\begin{align}
     \nabla \cdot  \bigg[\mu \left( \frac{g}{a_0} \right) \vec g \bigg] = -4\pi G \rho,
    \label{EQ_MOND}
\end{align}
where $\vec g$ is the total gravitational field, and $\rho = \rho_i + \rho_e$ is the total mass density, containing a contribution sourcing the external field $\rho_e$, and the contribution describing the smaller masses $\rho_i$.
The function $\mu(x \gg 1) \to 1$ is the interpolating function that smoothly connects between the Newtonian and small acceleration regime.
Since the MOND equation is nonlinear in the field, the superposition principle does not hold, and the internal dynamics of subsystem cannot be decoupled from the external field \cite{MOND-Milogram1983,Lag4MOND-Bekenstein1984,EquiPri-Norton1985}.
In \Cref{APP_MOND} we provide concrete examples of MOND fields and illustrate the lack of superposition principle.
Nevertheless, when the external field dominates the internal one, the total field is approximated as
\begin{align}
 \vec{g} & \approx  \vec{g}_i + \vec g_e ;
    \label{EQ_G_SUP}
\\
g  &=  g_e \sqrt{1 + \left( \frac{g_i}{g_e} \right)^2 + 2 \frac{\vec g_i \cdot \vec g_e}{g_e^2}} 
\simeq g_e + \hat n_e \cdot \vec g_i,
\end{align}
where $\hat n_e = \vec g_e / g_e$ is the unit vector in the direction of the external field at the point of evaluation. This approximation is well justified in our setup, where the external field is sourced by the Earth $g_e \sim 10$ m/s$^2$, while the internal field $g_i \sim 10^{-10}$ m/s$^2$.
The interpolating function can therefore be expanded as
\begin{align}
    \mu\left(\frac{g}{a_0}\right) \approx \mu_e + \mu_e'  \, \frac{\hat n_e \cdot \vec g_i}{a_0},
\label{eq_linearexpansion}
\end{align}
where $\mu_e = \mu(x)|_{x=g_e / a_0}$, and $\mu_e'$ is the derivative over $x$ evaluated at $x=g_e / a_0$.
We now substitute \Cref{eq_linearexpansion,EQ_G_SUP} into \Cref{EQ_MOND}, together with $\rho = \rho_i + \rho_e$.
A part of the external-field dependence cancels due to
\begin{align}
     \nabla \cdot \left( \mu_e \, \vec g_e \right) = - 4 \pi G \rho_e.
    \label{APP_EQ_V_EXT}
\end{align}
Furthermore, it is legitimate to ignore the term $ \nabla \cdot \left( \mu_e' \, \frac{\hat n_e \cdot \vec{g}_i}{a_0} \, \vec g_i \right)$, since it is quadratic in the internal field and therefore suppressed by roughly ten orders of magnitude compared to the similar mixed term involving the external field.
The resulting equation, with only the internal mass distribution, becomes
\begin{align}
     \nabla \cdot \left( \mu_e \vec g_i + \mu_e' \, \frac{\hat n_e \cdot \vec{g}_i}{a_0} \, \vec g_e \right)
     =  - 4 \pi G \rho_i.
    \label{APP_EQ_GEGI}
\end{align}
For comparison with existing literature, we divide by $\mu_e$ and define the parameter $\lambda_e = \frac{\mu_e'}{\mu_e} \frac{g_e}{a_0}$, yielding
\begin{align}
  \nabla \cdot \big( \vec g_i + \lambda_e \, (\hat n_e \cdot \vec{g}_i) \hat n_e \big) & =  - \frac{ 4 \pi G \rho_i }{ \mu_e } . \label{EQ_L} 
 \end{align}
The second term on the left side describes the coupling between the internal dynamics and the external field and is responsible for EFE. This term depends on the projection of $\vec g_i$ onto $\vec g_e$, and explicitly introduces an anisotropy leading to violation of the equivalence principle.

The argument so far is general, and applies to arbitrary mass distribution provided that the resultant field is much smaller than the external field. 
In the following section, we investigate the special case where a point mass embedded in an uniform external field.

%%%%%%%%%%%%%%%%%%%%%%%%%%
\section{Uniform external field}
\label{SEC_UNIFORM}
%%%%%%%%%%%%%%%%%%%%%%%%%%

Consider a point mass $m$ located at the origin, embedded in a uniform external field directed along the negative $z$ axis, $\vec g_e = g_e (- \hat z)$, such that \Cref{EQ_L} simplifies to:
\begin{align}
 \nabla \cdot \left( \vec g_i + \lambda_e \, (\hat z \cdot \vec{g}_i) \hat z \right) 
=  - \frac{ 4 \pi G m \delta(\vec r) } { \mu_e },
\label{EQ_UNI_EFE}
\end{align}
Let us call the field under the divergence as $\vec f = \vec g_i + \lambda_e \, (\hat z \cdot \vec{g}_i) \hat z$. Since the system admits azimuthal symmetry and has a Dirac delta source, we seek solutions of the form
\begin{equation}
\vec f = -  \frac{G m}{r^2} \frac{f(\theta)}{\mu_e} \hat r,
\label{EQ_ANS}
\end{equation}
where $\hat r$ is the radial unit vector, and $\theta$ is the polar angle in the usual spherical coordinates. 
\Cref{EQ_UNI_EFE} with the ansatz in \Cref{EQ_ANS} gives the constraint:\footnote{In general, $\nabla \cdot \left( f(\theta) \frac{\hat{r}}{r^2} \right)=4\pi C \delta(\vec{r})$ requires $f(\theta)$ to satisfy $\int_0^{2\pi}d\theta \, \sin\theta \, f(\theta)=2C$, where $C$ is a dimensionless constant.}
$\int_0^\pi d \theta \, \sin \theta \, f(\theta) = 2$.
Comparing the definition of $\vec f$ with the ansatz in \Cref{EQ_ANS}, we get the internal field components as
\begin{align}
g_{ir} & =  - \frac{G m}{r^2} 
\frac{f(\theta)}{\mu_e(1 + \lambda_e)}
( 1 + \lambda_e \sin^2 \theta ) ,
\notag\\%%%%%%%%%%%%%%%%%%
g_{i\theta} &=  - \frac{G m}{r^2} 
\frac{f(\theta)}{\mu_e(1 + \lambda_e)}
 \lambda_e \sin \theta \cos\theta ,
\notag\\%%%%%%%%%%%%%%%%%%
g_{i\phi} &= 0 .
\label{EQ_GEN_FIELD}
\end{align}
A variety of functions $f(\theta)$ satisfy these requirements.
However, a unique solution emerges if we impose the requirement of vanishing curl, in analogy with Newtonian gravity for static mass distributions.

%%%%%%%%%%%%%%%%%%%%%%
\subsection{Vanishing curl}

The vanishing curl constraint on the field in \Cref{EQ_GEN_FIELD} yields the following first-order differential equation:
\begin{equation}
(1 + \lambda_e \sin^2 \theta) \dv{\theta} f(\theta) + 3 \lambda_e \sin \theta \cos\theta \, f(\theta) = 0 .
\end{equation}
The solution is given by
\begin{equation}
f(\theta) = \frac{1 + \lambda_e}{(1 + \lambda_e \sin^2 \theta)^{3/2}} ,  
\label{EQ_FT_BN}
\end{equation}
which leads to the result derived in Ref.~\cite{Lag4MOND-Bekenstein1984}, and studied in further detail by Blanchet and Novak \cite{EFE-Blanchet2011}.
The corresponding BN potential reads
\begin{align}
V_{\mathrm{BN}} = - \frac{\mathcal{G}_\theta m}{r} ;
\qquad
\mathcal{G}_{\theta} = \frac{G}{\mu_e \sqrt{1 + \lambda_e\sin^2\theta} }
,
\label{EQ_G_BN}
\end{align}
which resembles the Newtonian potential but with an anisotropic (angle-dependent) gravitational constant $\mathcal{G}_\theta$.

It predicts that the attraction is strongest in the parallel configuration of $\theta = 0$.
However, for $\theta = \pi/2$ the potential depends explicitly on $\lambda_e$, whereas \Cref{EQ_UNI_EFE} is independent of $\lambda_e$ in the equatorial-plane limit ($\hat z \cdot \vec g_i = 0$).
Also the field depends on $\lambda_e$, as seen from the radial component in \Cref{EQ_GEN_FIELD} with $f(\theta)$ given in \Cref{EQ_FT_BN}.
As demonstrated in \Cref{APP_RESCALING}, this dependence originates from scaling properties of the Dirac delta under rescaling of the $z$ coordinate.
Since physical predictions in the $xy$ plane should not be affected by rescaling the $z$ coordinate, this motivates going beyond the solutions constrained by a vanishing curl.

%%%%%%%%%%%%%%%%%%%%%%%%%%%%%%%%%%%%%
\subsection{Non-vanishing curl}

In Newtonian gravity, static fields are curl-free as a consequence of arising from the spatial gradient of the Newtonian gravitational potential. Similarly, field-based models of MOND typically have the gravitational field arising from the spatial gradient of one or more scalar fields, and so the resulting gravitational field is also curl-free. 

The equation \Cref{EQ_UNI_EFE} is somewhat more general in that it doesn't assume that $\vec{g}$ arises from the gradient of a scalar and so admits solutions where $\vec{g}$ possesses non-vanishing curl. 
For generality, we will consider this more exotic possibility and determine how predictions in the case of non-vanishing curl differ from the curl-free solution, particularly in the equatorial plane.

The simplest such solution is to set $f(\theta) = 1$.
It is apparent from \Cref{EQ_GEN_FIELD} that in the equatorial plane the field is indeed independent of $\lambda_e$.
A non-vanishing curl implies that there is no scalar potential in general - and we emphasize that this may be considered an exotic probability given that in General Relativity and known extensions of General Relativity that yield MOND-like phenomenology (for example \cite{TeVeS-Bekenstein2004} and \cite{RMOND-Zlosnik2021}) have $\vec{g}$ arising from spatial gradients of scalar potentials -
but since we are only interested in parallel and perpendicular configurations ($\theta = 0$ and $\pi/2$), an effective angle-dependent gravitational coupling still emerges.
In both cases the polar component vanishes, and the radial field reduces to an effective Newtonian form with modified gravitational constants
\begin{align}
\mathcal{G}_{\parallel}^\times = \frac{G}{\mu_e (1 + \lambda_e)} ,
\qquad
\mathcal{G}_{\perp}^\times = \frac{G}{\mu_e} ,
\label{EQ_G_CURL}
\end{align}
where the "$\times$" superscript signifies a non-vanishing curl.
Thus, in this formulation the attraction is stronger in the perpendicular configuration, and the actual magnitudes are different.
We now turn to the resulting dynamics and quantify the experimental precision required to resolve this difference.

\subsection{Spacetime trajectories and required precision}

The net effect of a strong uniform external field on internal dynamics is therefore an orientation-dependent Newton's constant as given in \Cref{EQ_G_BN,EQ_G_CURL}.
Once again, this motivates our proposal to verify the mutual attraction of two masses aligned parallel or perpendicular to the external field.
In principle, such an experiment can also discriminate between different EFE solutions, since they predict different force ratios. Let us denote $\mathcal{G}_{\parallel} = \mathcal{G}_0$ and $\mathcal{G}_{\perp} = \mathcal{G}_{\pi/2}$ appearing in the BN potential. Their ratio is determined by the interpolating function via $\lambda_e$, namely $\mathcal{G}_{\parallel} / \mathcal{G}_{\perp} = \sqrt{1 + \lambda_e}$.
A measurement of this ratio would therefore constrain the form of the MOND interpolating function.
For the non-vanishing curl solution, one instead finds $\mathcal{G}_{\parallel}^{\times} / \mathcal{G}_{\perp}^{\times} = 1/(1+\lambda_e)$, indeed a different ratio.

To make these differences quantitative, we consider a general family of interpolating functions $\mu(x,\alpha)$  \cite{MilgromSanders_2008}:
\begin{align}
\mu(x,\alpha > 0) = \frac{x}{(1 + x^\alpha)^{1/\alpha}}  .
\end{align}
They satisfy the requirements $\mu(x \ll 1,\alpha) \to x$ (in order to match galactic rotation curves), and $\mu(x \gg 1,\alpha) \to 1$ (to recover Newtonian dynamics in strong external gravity).
The so-called ``standard'' MOND form is recovered for $\alpha=2$ \cite{Kent_1987,MOND-Milgrom1988} 
and the ``simple'' form for $\alpha=1$ \cite{MONDinMilkyWay-Famaey2005}, while other values are also allowed.
For a uniform external field comparable to Earth’s gravity, $g/a_0 \approx 10^{11} \gg 1$, one finds $\lambda_e = (1 + (g_e/a_0)^\alpha)^{-1} \approx 10^{-11 \alpha}$. 
Consequently, $\mathcal{G}_{\parallel} / \mathcal{G}_{\perp} \approx 1 + 10^{-11 \alpha}$ and $\mathcal{G}_{\parallel}^{\times} / \mathcal{G}_{\perp}^{\times} \approx 1 - 10^{-11 \alpha}$.
For typical values $\alpha \sim 1$, the deviation appears only at the 11th decimal place or beyond.
Such effects are clearly challenging experimentally, but we note that smaller values of $\alpha$ still remain interesting, especially as tests of validity of the equivalence principle, and that the experiment proposed here would place a concrete bound on $\alpha$.
%%%%%%%%%%%%%%%%%%%%%%%%%%%%%%%%%%%%%
\begin{figure*}[!t]
    \centering
    \includegraphics[width=0.9\linewidth]{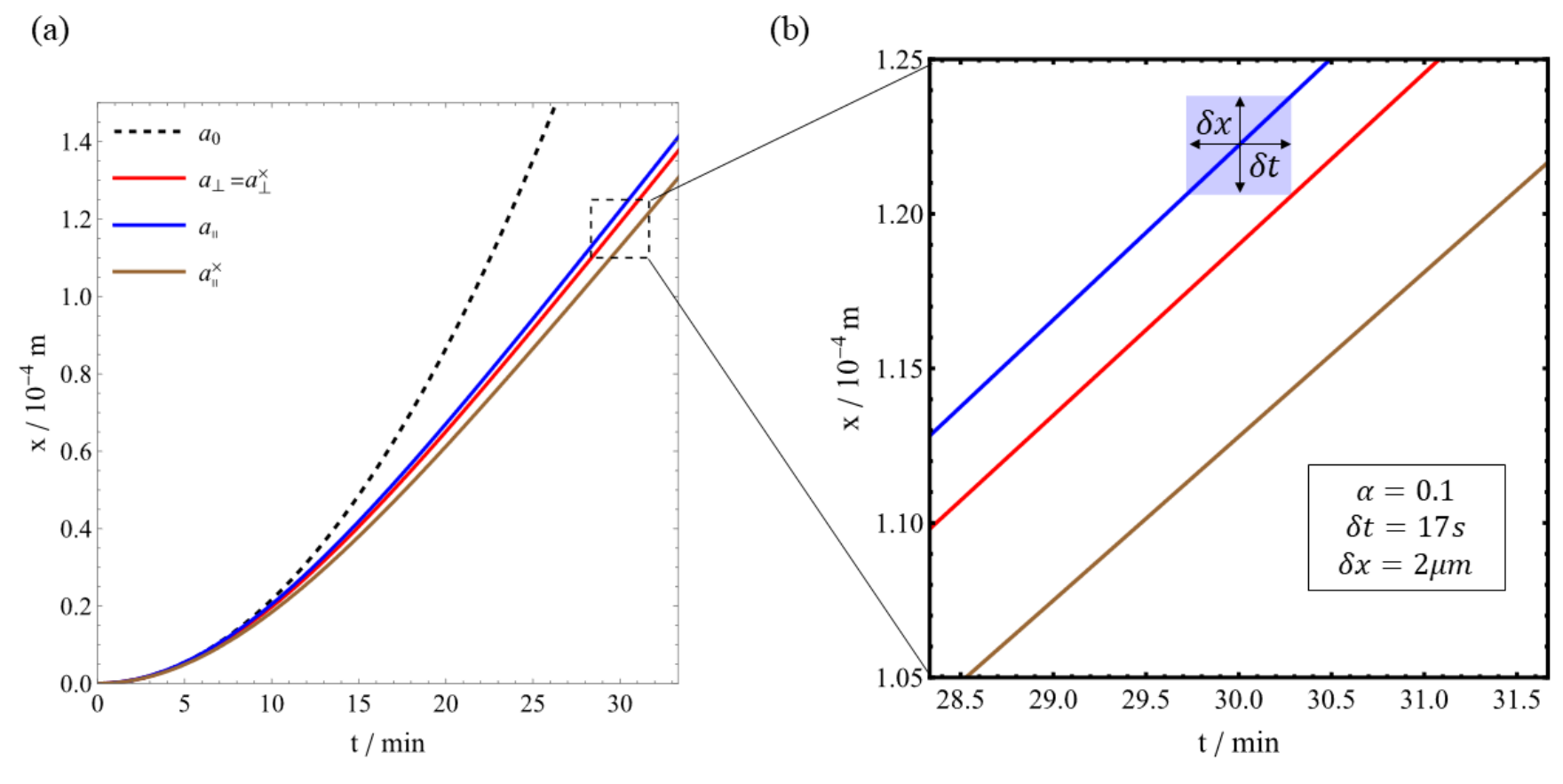}
    \caption{Spacetime trajectories of Platinum spheres. Each of the two spheres has radius $215 \,\mu$m, and their final surface-to-surface separation is $274 \,\mu$m. They are embedded in a uniform external field of Earth's strength. In panel (a), the dashed curve corresponds to uniformly accelerated motion with acceleration $a_0$. This approximation remains accurate for evolution times up to approximately one minute. The solid curves show the MOND trajectories for an interpolating function with $\alpha = 0.1$ and the exact value of $\lambda_e$. The red and blue curves correspond to the Blanchet–Novak solution, \Cref{EQ_G_BN}, in the parallel and perpendicular configurations, respectively, while the red and brown curves show the analogous trajectories for the non-vanishing curl solution, \Cref{EQ_G_CURL}.
    Newtonian mechanics follows the red curve in both configurations.
    Panel (b) displays a magnified view of the trajectories near $t = 30$ minutes. Resolving the predicted differences experimentally requires temporal precision better than $\delta t < 17$ s and spatial precision better than $\delta x < 2 \,\mu$m. Testing the so-called simple interpolating function is substantially more demanding, requiring both resolutions approximately seven orders of magnitude better than quoted here.}
    \label{FIG_TRAJECTORY}
\end{figure*}
%%%%%%%%%%%%%%%%%%%%%%%%%%%%%%%%%%%%%
Even in the context of MOND parameters, experiments testing smaller values of $\alpha$ may be of interest for several reasons. At present, MOND phenomenology exhibits a degree of tension between different observational constraints. On the one hand, fits to galactic rotation curves tend to favor interpolating functions close to the “simple” form, corresponding to $\alpha = 1$~\cite{MONDinMilkyWay-Famaey2005,Desmond2024}. We have also verified that a subset of the SPARC~\cite{Sparc} (Spitzer Photometry and Accurate Rotation Curves) sample studied in Ref.~\cite{Wang_2021} can be well fitted with $\alpha = 0.7$, using the standard value of $a_0$ and allowing for adjustments in the mass-to-light ratio.
On the other hand, constraints from the Cassini measurements of the Sun–Saturn system appear to prefer significantly larger values of $\alpha$. Theoretical estimates of EFE in the Solar System require $\alpha \gtrsim 5$ to remain consistent with the observational bounds on the quadrupole parameter~\cite{EFE-Blanchet2011,Desmond2024}. 
We note, however, that these analyses generally do not account for solutions with non-zero curl fields.
This leaves open the question of whether such solutions could alleviate the apparent tension and permit smaller values of $\alpha$. 
This question cannot be addressed with the solutions derived above because the regime considered in the present work is qualitatively different. In particular, the Solar System is embedded in the external gravitational field of the Milky Way, estimated to be approximately $1.6 \, a_0$, while the internal gravitational field of the Sun at Saturn’s orbit exceeds this by roughly five orders of magnitude. In contrast, the present solutions probe a regime in which the internal field is much weaker than the external field, providing a complementary and experimentally distinct test of EFE.

The effective gravitational constants derived above differ from the Newtonian constant $G$. If MOND is correct, standard Cavendish experiments effectively measure the gravitational constant in the orthogonal orientation.
Accordingly, in the numerical estimates below we set $\mathcal{G}_{\perp} = \mathcal{G}_{\perp}^{\times} = 6.67408 \times 10^{-11} \, \text{Nm}^2/\text{kg}^2$.

We now determine the spatial and temporal measurement precision required to exclude values of $\alpha$ below a given threshold. For an illustration, \Cref{FIG_TRAJECTORY} gives the spacetime trajectory of each Platinum sphere of radius $215 \,\mu$m and final surface-to-surface separation of $274 \,\mu$m (only due to internal forces). The interpolating function is taken with $\alpha = 0.1$. According to \Cref{EQ_T_X} with effective gravitational constant, the difference in travel time to reach a fixed displacement in the two configurations is $| t_{\perp} - t_{\parallel} | = t |\sqrt{\mathcal{G}_{\parallel} / \mathcal{G}_{\perp}} - 1| \simeq t \lambda_e / 4$ for both EFE solutions considered, where $t = t_{\parallel}$ denotes the time at which the displacement is evaluated. The temporal resolution of a probe must be smaller than half of this difference:
\begin{equation}
\delta t < \frac{t \lambda_e}{8}.
\label{EQ_T_RES}
\end{equation}
This requirement is relatively mild. For an evolution time of $t = 1800$ seconds (half an hour), \Cref{EQ_T_RES} yields $\delta t < 17$ s for $\alpha = 0.1$, and $\delta t < 2$ ns for the simple interpolating function with $\alpha = 1$.  
Only the standard function, $\alpha = 2$, would require an extremely demanding temporal resolution of $\delta t < 2 \times 10^{-20}$ s. We emphasize that \Cref{FIG_TRAJECTORY} gives the plot of the internal dynamics, i.e. as would be observed by levitating the masses with a non-gravitational force equal and opposite to the uniform external gravity or possibly in space-based laboratory.

We now estimate the required spatial precision. Determining the displacement at time $t$ requires solving \Cref{EQ_T_X}, which is transcendental and does not admit a closed-form solution. Yet, it can be readily solved numerically and therefore let $x_{\parallel}$ denote the solution of $t = f(x_{\parallel})$, where function $f$ is the right-hand side of \Cref{EQ_T_X} with the effective gravitational constant for the parallel configuration. For the perpendicular case, the corresponding displacement $x_{\perp}$ satisfies $t = \sqrt{\mathcal{G}_{\parallel} / \mathcal{G}_{\perp}} f(x_{\perp})$, so that
\begin{equation}
f(x_{\perp}) = t \sqrt{\mathcal{G}_{\perp} / \mathcal{G}_{\parallel}} 
\simeq t - t \lambda_e/4,
\label{EQ_F_X_PERP}
\end{equation}
i.e., $x_{\perp}$ is the solution to the same equation, but for a suitably earlier time.\footnote{The time is suitably later for the solution with non-vanishing curl as $\sqrt{ \mathcal{G}_{\perp}^{\times} / \mathcal{G}_{\parallel}^{\times} } \simeq 1 + \lambda_e / 2$. }
To estimate the displacement difference, we expand $f(x)$ around $x_{\parallel}$,
\begin{equation}
f(x) \simeq f(x_{\parallel}) + (x - x_{\parallel}) f'_{\parallel},
\label{EQ_EXP_F}
\end{equation}
where $f'_{\parallel} = \dv{f}{x}|_{x = x_{\parallel}}$.
Since $f(x_{\parallel}) = t$, comparison of \Cref{EQ_EXP_F} for $x = x_{\perp}$ with \Cref{EQ_F_X_PERP} gives $ | x_{\parallel} - x_{\perp} | \simeq t \lambda_e / 4 f'_{\parallel}$. Thus, the spatial resolution requirement is
\begin{equation}
\delta x < \frac{t \lambda_e}{8 f'_{\parallel}}.
\end{equation}
Unlike the temporal case, the spatial sensitivity is further suppressed by the derivative $f'_{\parallel}$. For the parameters in \Cref{FIG_TRAJECTORY}, one finds $f'_{\parallel} \approx 6.8 \times 10^6$, which effectively reduces the relevant scale by about seven orders of magnitude. Concretely, to test the MOND model with $\alpha = 0.1$ one requires $\delta x < 2 \,\mu$m. 
For the simple interpolating function with $\alpha = 1$, the requirement tightens to $\delta x < 3 \times 10^{-16}$ m, while for the standard form with $\alpha = 2$ it becomes $\delta x < 3 \times 10^{-27}$ m. Although the latter cases are extremely challenging, it is noteworthy that optomechanical techniques are already approaching the regime relevant for the simple function \cite{CoolGroundState-Teufel2011,LaserCoolNanoOsc-Aspelmeyer2011}. 

A small improvement in spatial resolution occurs for larger masses due to the scaling of $f'_{\parallel}$ with mass. Larger masses must be placed farther apart to maintain acceleration below $a_0$, so the resulting displacements remain much smaller than the separation. In this regime, the uniform-acceleration approximation becomes accurate.
For instance, $t_{\parallel}^2 = 2 x L_f^2 / \mathcal{G}_{\parallel} m$ and $t_{\perp}^2 = \big( 2 x L_f^2 / \mathcal{G}_{\parallel} m \big) \sqrt{1 + \lambda_e}$. 
The corresponding temporal resolution is still given by \Cref{EQ_T_RES}. 
For spatial resolution, one finds, $x_{\parallel} =   \frac{1}{2} a_0 t^2$ and $x_{\perp} = \frac{1}{2} a_0 t^2 (1+\lambda_e)^{-1/2}$, leading to $\delta x < a_0 t^2 \lambda_e / 8$.
As an example, consider Platinum spheres of radius $10$ cm. 
The required separation becomes $L_f \approx 7$ m, while the displacement remains below the millimeter scale. For $t = 30$ min, one finds $\delta x < 4 \,\mu$m given $\alpha = 0.1$. Note that increasing the mass by nine orders of magnitude improves the spatial requirement by only a factor of two.
%%%%%%%%%%%%%%%%%%%%%%%%%%%%%%%%%%%%%
\begin{figure}[!t]
    \centering
    \includegraphics[width=\linewidth]{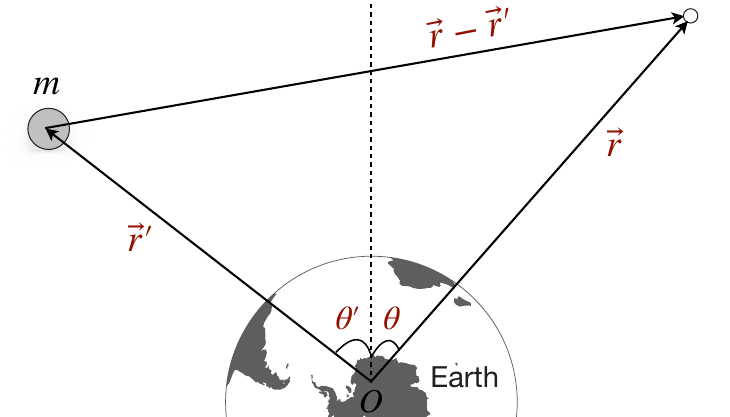}
    \caption{A point mass $m$ located at position $\vec r\,'$ in radial external field generated by the Earth. Presented variables are used in \Cref{EQ_GI_RAD} to write the internal gravitational field evaluated at position $\vec r$.}
    \label{FIG_RAD_1}
\end{figure}
%%%%%%%%%%%%%%%%%%%%%%%%%%%%%%%%%%%%%

%%%%%%%%%%%%%%%%%%%%%%%%%%%%%%%%%%%%%%%%%%%
\subsection{Region of uniform gravity}

The results so far hold for uniform external field.
Recall from \Cref{EQ_G_SUP} that in strong fields the superposition principle remains approximately true and hence the trajectories plotted in \Cref{FIG_TRAJECTORY} would be observed in a reference frame accelerating with the external field.
Assuming the Earth is a perfect sphere gives rise to radial gravitational field around it and the corresponding tidal effects.
Accordingly, there is no uniformly accelerating reference frame in which the effects of external gravity are completely removed.
We now estimate the region in Earth's gravity where its field could be considered uniform for practical purposes.
Since the measurements proposed here necessarily take long time, the particles should be balanced by a non-gravitational force to keep them from free falling (alternatively, the experiment could be performed in space-based environments). We assume this additional force is uniform in space, satisfies the superposition principle, and balances exactly the center of mass between the particles.
In this arrangement, the field is considered uniform if the changes in external gravity at the beads are smaller than the MOND acceleration $a_0$.

Consider the parallel configuration first, i.e. the masses are arranged on top of each other. For the estimation we take their center-of-mass to be located at the radius of the Earth, $R_{\oplus} = 6.317 \times 10^6$ m, and we call the separation between them $h_{\parallel}$.
Accordingly, the external gravity at the position of the beads is $g \simeq g_{\oplus}(1 \pm h_{\parallel}/R_{\oplus})$, with the stronger field for the particle closer to the center of the Earth, and where the acceleration of the center of mass is $g_{\oplus} = 9.81$ m/s$^2$. 
In the arrangement described above, the two particles effectively move away from each other with acceleration $g_{\oplus} h_{\parallel}/R_{\oplus}$.
This value should be smaller than $a_0$, bounding the radial displacement to
\begin{align}
    h_{\parallel} \simeq R_{\oplus} \frac{a_0}{g_{\oplus}} \simeq 80 \,\mu\text{m}.
\end{align}

In the perpendicular scenario, the external field at the beads is different both in magnitude and in direction. We call the horizontal separation as $h_{\perp}$ and again assume that the center of mass experiences acceleration $g_{\oplus}$ pointing downwards. The external field at the beads has magnitude $g \simeq g_{\oplus} (1 - h_{\perp}^2/2R_{\oplus}^2)$ and points towards the center of the Earth. Therefore, after balancing the center of mass, the particles are attracted along the symmetry axis with acceleration $\simeq g_{\oplus} h_{\perp} / 2 R_{\oplus}$, and they admit an upwards component $\simeq 3 g_{\oplus} h_{\perp}^2 / 8 R_{\oplus}^2$. The net trajectory is an arc above the symmetry axis. Since the upward component scales quadratically, the size restriction comes from the acceleration along the symmetry axis and reads:
\begin{align}
    h_{\perp} \simeq 2 R_{\oplus} \frac{a_0}{g_{\oplus}} \simeq 160 \,\mu\text{m}.
\end{align}

These bounds quantify the spatial region within which Earth's gravitational field may be treated as effectively uniform. Note that they are obtained from comparing changes in gravity to the critical acceleration $a_0$, and hence the volume obtained is the upper bound on the volume that should actually be used, where ideally one would like to ensure that the changes in external gravity are a fraction of $a_0$.
Yet, even the distances in this upper bound are smaller than the required surface-to-surface distance between the particles derived in \Cref{SEC_VDWAALS}. This motivates considering the EFE for radial fields.

%%%%%%%%%%%%%%%%%%%%%%%%%%%%%%%%%%%%%%%%%%%
\section{Radial external field}
\label{SEC_RADIAL}
%%%%%%%%%%%%%%%%%%%%%%%%%%%%%%%%%%%%%%%%%%%

%%%%%%%%%%%%%%%%%%%%%%%%%%%%%%%%%%%%%
\begin{figure}[!t]
    \centering
    \includegraphics[width=\linewidth]{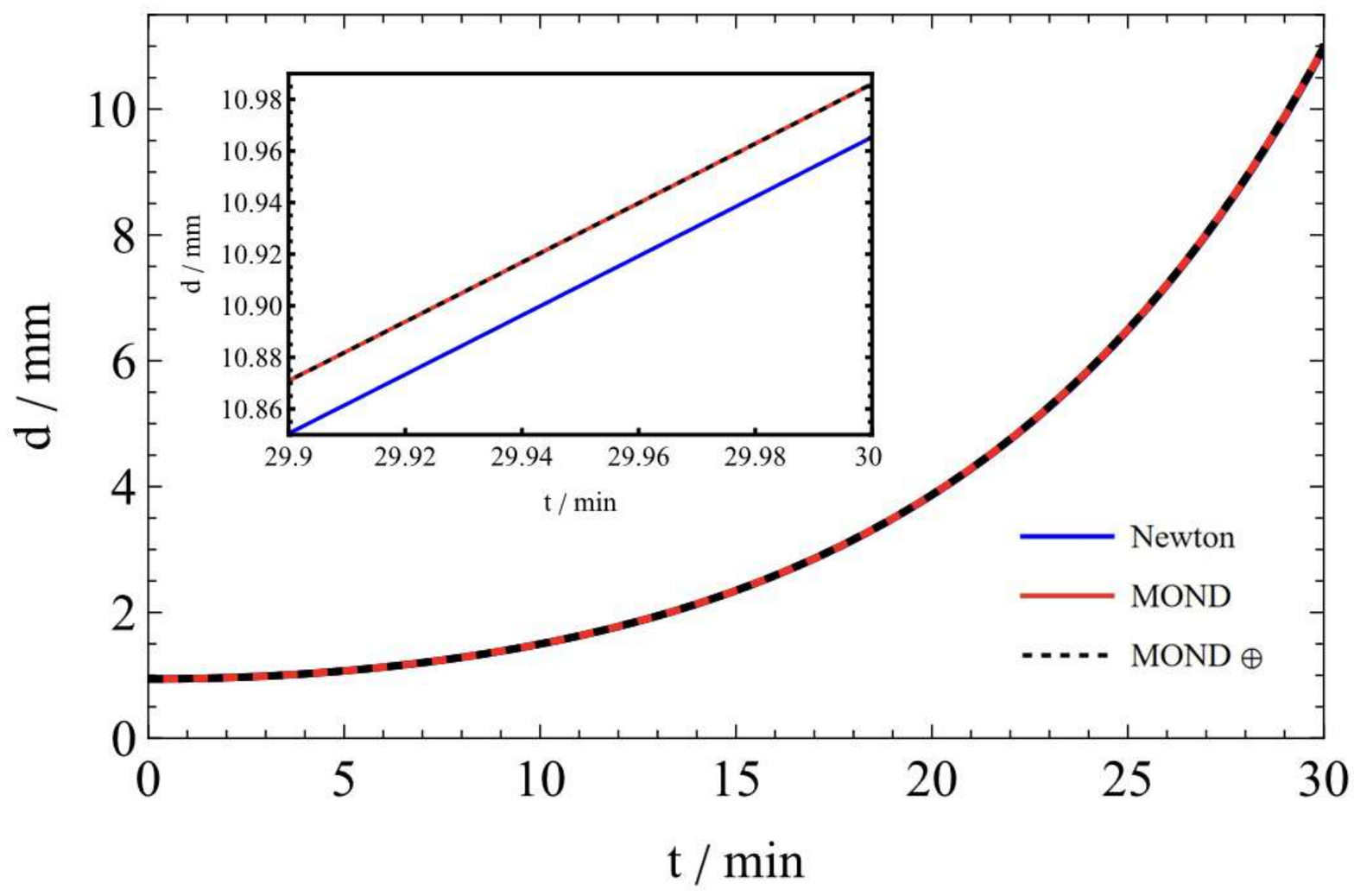}
    \caption{Evolution of the distance between the particles in the parallel configuration embedded in the radial external gravitational field and with their center of mass balanced by a non-gravitational force uniform in space. The tidal forces coming from external gravity pull the particles apart ten times stronger than their internal attraction, causing increasing separation. The curve denoted MOND $\oplus$ shows dynamics where interpolating function has a fixed value, given by the value at the center of mass. The non-uniformity in $\mu_e$ is practically undetectable. The inset magnifies the curves close to $30$ minute evolution  time and shows that the required spatial precision to distinguish Newtonian and MOND solution with curl is the same as for the solution with curl in the uniform external field.}
    \label{FIG_RAD_PAR}
\end{figure}
%%%%%%%%%%%%%%%%%%%%%%%%%%%%%%%%%%%%%

Consider the configuration shown in \Cref{FIG_RAD_1}. A point mass $m$ is located at position $\vec r\,' = (r',\theta')$, measured from the center of the spherical Earth. Applying \Cref{EQ_L} to a radial external field and the point mass shows that the internal field $\vec g_i$ satisfies
\begin{equation}
(1 + \lambda_e) g_{ir} \hat r + g_{i\theta} \hat \theta = -G \frac{m}{\mu_e} \frac{\vec r - \vec r\,'}{|\vec r - \vec r\,'|^3},
\end{equation}
where the field is evaluated at position $\vec r = (r,\theta)$. The quantities $\mu_e$ and $\lambda_e$ are also evaluated locally, since they depend on the external field magnitude $g_e$ that now varies from point to point.
Again, there are many possible solutions to the internal field and for simplicity we choose one with non-zero curl.
Comparing components on both sides gives
\begin{align}
g_{ir} & = - \frac{G m}{\mu_e} \frac{1}{1+\lambda_e} \frac{r - r' \cos(\theta + \theta')}{ \left( r^2 + r'^2 - 2 r r' \cos(\theta + \theta') \right)^{3/2}},
\notag\\%%%%%%%%%%%%%%%%%%%%%
g_{i \theta} & = - \frac{G m}{\mu_e} \frac{r' \sin(\theta + \theta')}{\left( r^2 + r'^2 - 2 r r' \cos(\theta + \theta') \right)^{3/2}}.
\label{EQ_GI_RAD}
\end{align}
The parallel configuration corresponds to $\theta = \theta' = 0$, in which case the internal field is purely radial, with effective Newton constant:
\begin{equation}
\mathcal{G}_{\parallel}^r = \frac{G}{\mu_e (1 + \lambda_e)}.
\end{equation}
This has the same form as the non-vanishing curl solution for the uniform external field, although here $\mu_e$ and $\lambda_e$ depend explicitly on position.
Therefore, the forces acting on the two masses are not equal in magnitude, i.e., Newton's third law is violated. 
Denoting the acceleration of middle point between the particles (COM) as $g_{\oplus}$, 
the parameters of the model at the location of the beads, expanded to the first order, read
\begin{align}
g_{e} & =  g_{\oplus} \left( 1 \pm \frac{h}{R_{\oplus}} \right), 
\notag\\%%%%%%%%%%%%%%%%%%
\mu_e & =  \mu_{\oplus} \pm \mu_{\oplus} ' \frac{g_{\oplus}}{a_0} \frac{h}{R_{\oplus}},
\notag\\%%%%%%%%%%%%%%%%%%
\lambda_e & =  \frac{1}{1+\left( \dfrac{g_{\oplus}}{a_0} \right)^{\alpha} \left( 1 \pm \dfrac{h}{R_{\oplus}} \right)^{\alpha}} ,
\end{align}
where $h$ is the separation between the masses,
$R_\oplus$ is the radial distance of the COM from Earth's center, and
$\lambda_\oplus$ and $\mu_\oplus$ denote the values of $\lambda_e$ and $\mu_e$ evaluated at the location of the COM.
For the parameters used in \Cref{FIG_TRAJECTORY}, the correction factor $h/R_{\oplus} \sim 10^{-9}$, implying that $\lambda_e \simeq \lambda_{\oplus}$ and $\mu_e \simeq \mu_{\oplus}$ to at least eleven decimal places for $\alpha = 0.1$ (with even better agreement for larger $\alpha$).
Therefore, if radial external field was balanced everywhere in space by a non-gravitational force, the trajectory of the beads is practically the same as for the uniform external field.
If one balances only the center of mass between the particles, with a force that is uniform in space, \Cref{FIG_RAD_PAR} shows how the distance between the particles evolves in Newtonian mechanics and in MOND model.
%%%%%%%%%%%%%%%%%%%%%%%%%%%%%%%%%%%%%
\begin{figure}[!t]
    \centering
    \includegraphics[width=0.8\linewidth]{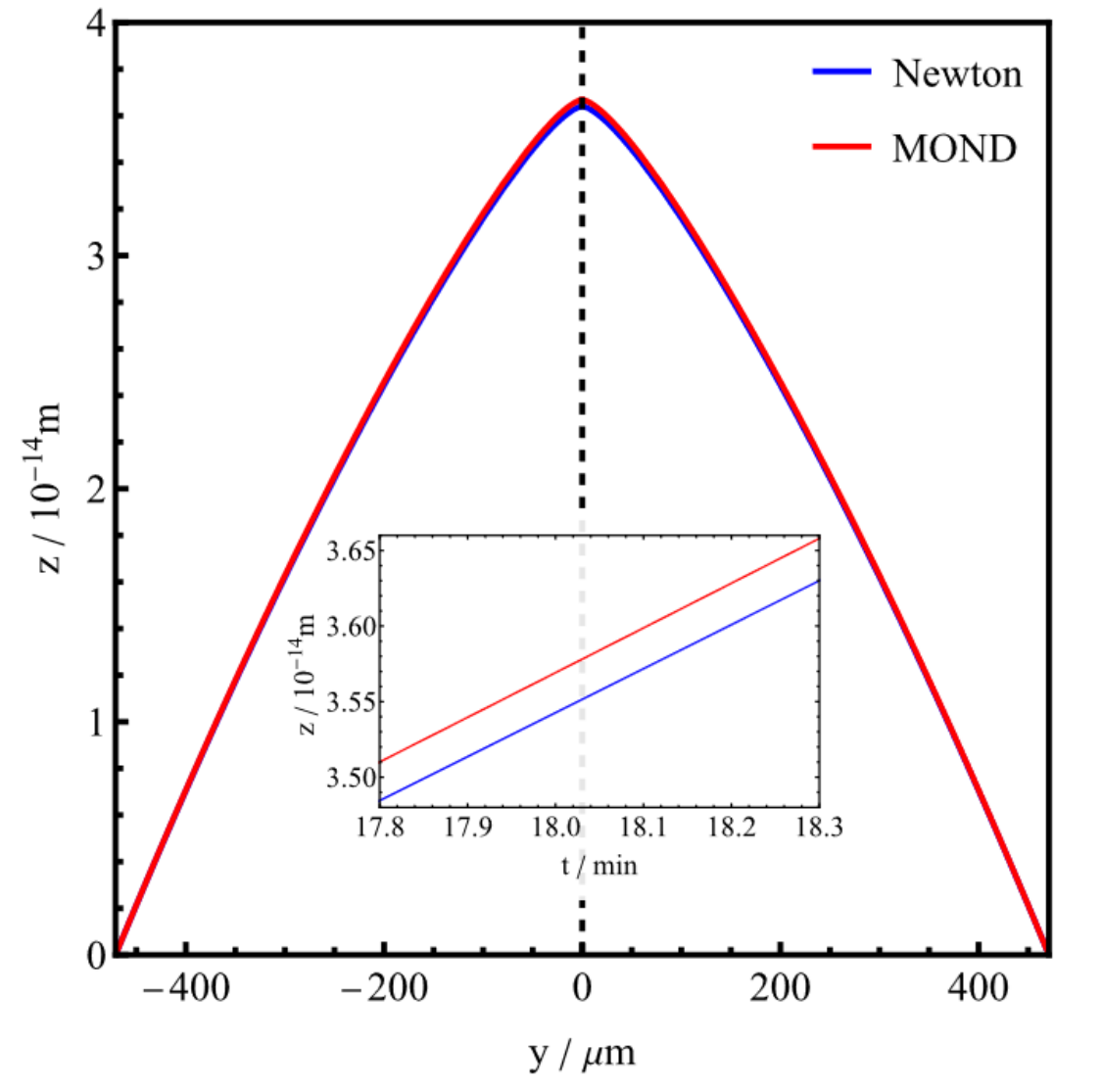}
    \caption{Trajectory of the two masses in the orthogonal configuration under a radial external field, and with their center of mass balanced by a non-gravitational force uniform in space. The parameters are identical to those used in \Cref{FIG_TRAJECTORY}. For illustration, the trajectories are plotted up to the point of collision, although the condition that the internal acceleration remains below $a_0$ is satisfied only for displacements along the $y$ axis up to approximately $120 \, \mu$m (the net acceleration is at most $6 a_0$ in this reference frame). The masses are initially placed at $z = 0$. The arc in Newtonian mechanics is the result of tidal forces. Remarkably, the MOND arc is present even when the radial external field is balanced at every point in space.}
    \label{FIG_RAD_TRAJ}
\end{figure}
%%%%%%%%%%%%%%%%%%%%%%%%%%%%%%%%%%%%%

The perpendicular configuration exhibits qualitatively different behavior. In this case, the particles' symmetry axis is along the direction orthogonal to the dotted line in \Cref{FIG_RAD_1}, i.e. $r' = r$ and $\theta' = \theta$ at all times.
Let us denote $(y,z)$ the Cartesian coordinate system with origin at the center of mass between the particles, $z$ axis along the dotted line pointing upwards, and $y$ axis along the symmetry line pointing to the right.
It is instructive to write the internal field using the mixture of these coordinates and the spherical ones:
\begin{eqnarray}
g_{iy} & = & - \frac{G m}{\mu_e (2 y)^2} \left( \frac{ \sin^2 \theta}{1 + \lambda_e} + \cos^2 \theta \right), \nonumber \\
g_{iz} & = & \frac{G m}{\mu_e (2 y)^2}  \frac{\lambda_e \sin \theta \cos \theta}{1 + \lambda_e}.
\end{eqnarray}
It is now evident that for $\lambda_e = 0$, the internal field would point along the symmetry axis connecting the particles, and in fact it would have the Newtonian form (note that $2y$ is the distance between the particles along the symmetry axis).
Since $\lambda_e > 0$ in the MOND model, the net internal force tilts upward relative to the symmetry axis. Consequently, even if the external field was balanced everywhere in space by forces of non-gravitational nature, the particles would move toward each other along an arc above the symmetry axis.
This provides a vivid illustration of the violation of Newton's third law in the exotic case where $\vec{g}$ may possess a non-vanishing curl. The internal mutual forces acting on the particles (after balancing the radial external field everywhere) are not equal and opposite, but rather mirror reflected with respect to the $z$ direction. 
If one balances only the
center of mass between the particles, with a force uniform
in space, the resulting two-dimensional arcs are shown in 
\Cref{FIG_RAD_TRAJ}. We also verified analytically that the order of magnitude of the displacement along the $z$ axis agrees with   numerical results.

%%%%%%%%%%%%%%%%%%%%%%%%%%%%%%%%%%
\section{Conclusions}
\label{SEC_CONCLUSIONS}
%%%%%%%%%%%%%%%%%%%%%%%%%%%%%%%%%%

We have proposed precision experiments that compare the internal gravitational acceleration between two spherical masses with symmetry axis oriented either parallel or orthogonal to a strong external gravitational field. Although the calculations were carried out within the framework of MOND, the proposed setup allows an independent test of gravity itself, in particular of the strong equivalence principle. It would therefore be valuable to perform such an experiment even at precision that is insufficient to probe MOND.

We first analyzed the internal two-body dynamics in the absence of an external field, accounting for a finite size of the spheres, surface interactions, air drag, and time constraints. This led to a concrete benchmark configuration involving Platinum spheres of radius $215 \,\mu$m with the final surface-to-surface separation of $274 \,\mu$m (another relevant configuration is more macroscopic: with radius 2 cm and final inter-surface separation 60 cm high-vacuum conditions are not necessary).
We then studied the external field effect in a uniform background field and derived MOND internal dynamics with non-zero curl of the field as well as the spatial and temporal resolution required to distinguish the parallel and orthogonal configurations. The dominant limitations arise from spatial resolution and long experimental time.
Finally, we extended the analysis to a radial external field and found an unexpected feature in the configuration where the symmetry axis is orthogonal to the external field at the center of mass. The two masses do not attract each other strictly along their connecting line. Instead, the internal dynamics acquires a small transverse component that effectively drives the particles along an arc, even when radial external field is balanced everywhere in space.

%%%%%%%%%%%%%%%%%%%%%%%%%%%%%%%%%%
\begin{acknowledgments}
This research is supported by
Xiamen University Malaysia (Grant No. XMUMRF/2022-C10/IPHY/0002 and Grant No. XMUMRF/2021-C8/IPHY/0001),
and Science and Engineering Research Board, Govt.~of India (Grant No. CRG/2022/009359).
A.K. was partially/fully supported by the Helen Diller Quantum Center at the Technion (Grant No. 86632417).
We acknowledge the National Supercomputing Mission (NSM) for providing computing resources of ‘PARAM Ganga’ at IIT Roorkee, which is implemented by C-DAC and supported by MeitY and DST, Govt. of India.
The authors warmly thank Borivoje Dakić for stimulating discussions.
\end{acknowledgments}
%%%%%%%%%%%%%%%%%%%%%%%%%%%%%%%%%%

%%%%%%%%%%%%%%%%%%%%%%%%%%%%%%%%%%%%
\appendix 
\crefalias{section}{appendix}
\Crefname{appendix}{Appendix}{Appendices}
%%%%%%%%%%%%%%%%%%%%%%%%%%%%%%%%%%%%

%%%%%%%%%%%%%%%%%%%%%%%%%%%%%%%%%%%%
\section{Comparison of MOND and Newtonian fields, and violation of the superposition principle}
\label{APP_MOND}
%%%%%%%%%%%%%%%%%%%%%%%%%%%%%%%%%%%%

Consider the mass distribution $\rho$, sourcing the potential $\Phi$ and the field $\vec g_M = -  \nabla \Phi$. The basic equation of the MOND model reads:
\begin{align}
 \nabla \cdot \bigg[ \mu \left( \frac{g_M}{a_0} \right) \vec g_M \bigg] = -4\pi G \rho.
    \label{APP_EQ_MOND}
\end{align}
While this nonlinear Poisson-like equation is fundamentally different from the Newtonian mechanics, it admits a form of Gauss' law obtained by integrating over a volume:
\begin{align}
    \oint \mu \left( \frac{g_M}{a_0} \right) \vec g_M \cdot d \vec a = -4\pi G m,
    \label{APP_EQ_GAUSS}
\end{align}
where $m$ is the mass enclosed within the volume.
We start with a comparison of Newtonian and MOND fields in symmetric configurations.

For a fixed symmetry, say spherical, cylindrical, or plane mass distributions, it is well known how to use Gauss' law to obtain Newtonian fields $\vec g_N$.
Note that \Cref{APP_EQ_GAUSS} differs from the usual Gauss' law in Newtonian mechanics by the presence of the interpolating function inside the flux. Therefore, in all such cases the relation between the Newtonian and MOND fields reads:
\begin{align}
\mu \left( \frac{g_M}{a_0} \right) \vec g_M = \vec g_N.
\label{APP_EQ_MN}
\end{align}
The MOND fields follow the direction of the Newtonian fields, but with a modified magnitude.
In the deep MOND regime, where $\mu(x) \to x$, \Cref{APP_EQ_MN} becomes quadratic in the magnitude with the solution:
\begin{align}
g_M = \sqrt{g_N a_0}.
\label{EQ_GMGN}
\end{align}

Let us now demonstrate explicitly how superposition principle is violated in the MOND model. Recall that in Newtonian mechanics this refers to the situation where a test particle experiences gravity from multiple sources and demands that the total field is the sum of the fields from individual sources.
For an illustration consider two parallel identical infinite planes, with surface density $\sigma_1$ and $\sigma_2$. By symmetry, the MOND field is perpendicular to the planes and on top of both of them it is given by the formula \Cref{EQ_GMGN} with $\sigma = \sigma_1 + \sigma_2$.
Therefore, the magnitude of the total field is not a linear sum but a Pythagoras-like sum of the MOND fields generated by the two surfaces individually: $g_{12}^2 = g_1^2 + g_2^2$. The same argument holds for any configuration where the symmetry of the combined system is the same as the symmetry of systems that are being combined.
For example, the Pythagorean-like formula would still be correct for a sphere inside a sphere, but it would not be correct for two point masses next to each other, as here the individual masses admit spherical symmetry but the combined system admits cylindrical symmetry.

%%%%%%%%%%%%%%%%%%%%%%%%%%%%%%%%%%%%
\section{Plane limit of the Blanchet-Novak potential}
\label{APP_RESCALING}
%%%%%%%%%%%%%%%%%%%%%%%%%%%%%%%%%%%%

Here we show that the origin of the dependence on $\lambda_e$ of the potential derived in Ref. \cite{Lag4MOND-Bekenstein1984,EFE-Blanchet2011} is in the scaling property of the Dirac delta.
Consider a point mass in a uniform external field.
Let us define the internal potential as $\vec g_i = -  \nabla V_{\mathrm{BN}}$.
\Cref{EQ_UNI_EFE} of the main text rewritten in terms of the potential is:
\begin{align}
    \qty( \nabla^2 + \lambda_e \partial^2_z ) {V_{\mathrm{BN}}} =  \frac{ 4 \pi G m \delta(\vec r) }{ \mu_e } .
    \label{APP_EQ_ANISO}
\end{align}
The following change of variables:
\begin{align}
(x',y',z') = \qty( x, y, \frac{z}{\sqrt{1 + \lambda_e}} ) ,
\end{align}
transforms the left side of \Cref{APP_EQ_ANISO} to an isotropic Laplacian, and the delta on the right is rescaled as:
\begin{align}
    \delta(\vec{r}\,') = \frac{\delta(\vec{r}\,)}{\sqrt{1+\lambda_e}} \, .
    \label{EQ_D_SCALE}
\end{align}
In this way the problem is reduced to the usual Poisson equation, admitting a solution:
\begin{align}
    V_{\mathrm{BN}}(r') = -\frac{GM}{r'}  \frac{1}{\mu_e \sqrt{1 + \lambda_e}}.
\end{align}
In the spherical coordinates one finds:
\begin{align}
    V_{\mathrm{BN}}(r, \theta) = -\frac{GM}{r} \frac{1}{\mu_e \sqrt{1+\lambda_e\sin^2{\theta}}},
\end{align}
where $\theta$ is measured relative to the direction of the external field.
For $\theta = \pi/2$ the potential depends on $\lambda_e$, and the origin of this dependence lies in the scaling of the $z$ coordinate, as performed in \Cref{EQ_D_SCALE}.

%%%%%%%%%%%%%%%%%%%%%%%%%%%%%%%%%%%%
%\bibliographystyle{apsrev4-2}
%\bibliography{cite.bib}
%%%%%%%%%%%%%%%%%%%%%%%%%%%%%%%%%%%%

%apsrev4-2.bst 2019-01-14 (MD) hand-edited version of apsrev4-1.bst
%Control: key (0)
%Control: author (72) initials jnrlst
%Control: editor formatted (1) identically to author
%Control: production of article title (-1) disabled
%Control: page (0) single
%Control: year (1) truncated
%Control: production of eprint (0) enabled
%

%%%%%%%%%%%%%%%%%%%%%%%%%%%%%%%%%%%%
\end{document}